# Your clean graphene is still not clean

*Ondrej Dyck[1], Aisha Okmi[2,3], Kai Xiao[1], Sidong Lei[4,5], Andrew R. Lupini[1], Stephen Jesse[1]*

[1] *Center for Nanophase Materials Sciences, Oak Ridge National Laboratory, Oak Ridge, TN, USA*
[2] *Department of Physical Sciences, Physics Division, College of Science, Jazan University, Kingdom of Saudi Arabia*
[3] *Department of Physics and Astronomy, Georgia State University, Atlanta, GA 30303, USA.*
[4] *NanoScience Technology Center, University of Central Florida, Orlanda, FL 32826*
[5] *Department of Materials Science and Engineering, University of Central Florida, Orlando, FL 32816*

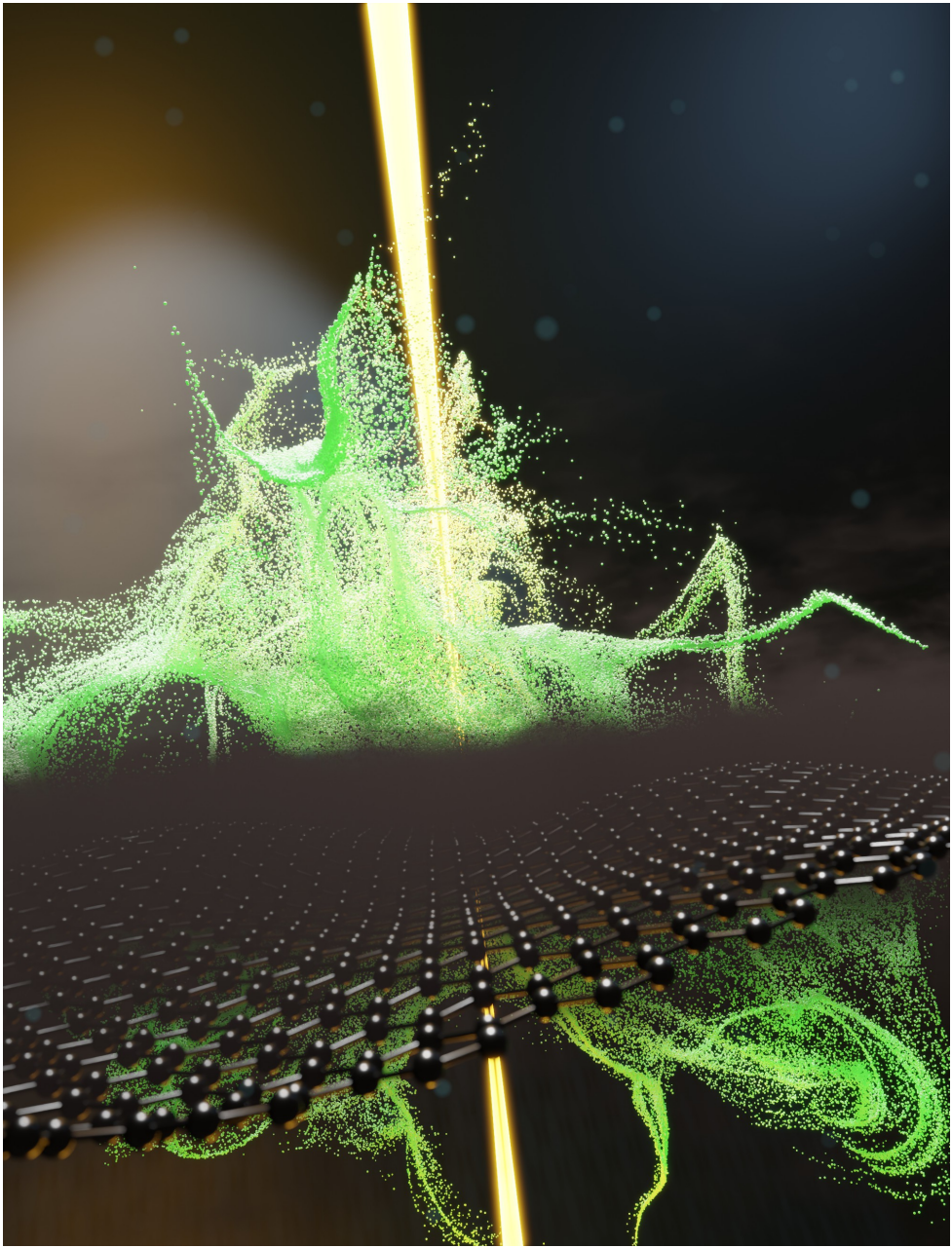

**Abstract**

Researchers working with thin samples, such as monolayer graphene, are consistently struggling against contamination. Indeed, the problem of hydrocarbon contamination has been known from the earliest days of electron microscopy and efforts to reduce this problem are ubiquitous to almost all high-vacuum experiments. Accurate knowledge of the behavior of such contamination is essential for electron beam (e-beam) based atomic fabrication, where we aspire to select and control matter on an atom-by-atom basis. Here, we take up the vexing question of hydrocarbon contamination on graphene. Image intensity is used to directly reveal the presence of diffusing hydrocarbons on ostensibly clean graphene. These diffusing hydrocarbons have been previously inferred but not directly observed. Surprising dynamic variations of the concentration of these hydrocarbons impels questions about their origin. Here we present some possible explanations and draw some tentative conclusions. This work updates the conceptual model of 'clean graphene' and offers refinements to the description of e-beam induced hydrocarbon deposition.

**Introduction**

There are many studies in the literature advancing a variety of strategies for cleaning graphene, providing contamination-free imaging in the (scanning) transmission electron microscope ((S)TEM).[1–16] One in particular asks in the title 'How clean can it be?'[10] and shows images of what appears to be atomically clean graphene over areas extending up to thousands of square nanometers. This result is not isolated, numerous other authors have also presented atomically resolved images of large areas of what appears to be clean graphene.[3,9,11,12,15,16] The unstated implication is that, if we have atomically resolved images of graphene, then there is no longer any doubt that it is perfectly clean. But there remain lingering questions about cleanliness even when atomically resolved images of it can be acquired.

In recent work, Postl et al[17] pointed out that the observed damage rate for graphene was in disagreement with theoretical predictions. To account for this discrepancy, the authors make the inference that there must be spontaneous healing, which implies that carbon atoms are coming from somewhere, even if they cannot directly be observed. The authors present a quantitative model that can be used to account for the discrepancy by allowing for the presence of diffusing carbon atoms on the sample surface. Although the authors note there might be "some effects missing from our model", it is nevertheless highly suggestive that invisible surface carbon atoms or hydrocarbons play a non-trivial role in the behavior of graphene.

In an earlier work, Zan et al[18] showed what appear to be clean areas of graphene with holes that spontaneously heal. These are not simply single vacancies. In the case of single vacancies, we found that rapid vacancy diffusion could contribute to the apparent healing[19] and that this may be one of the missing effects mentioned by Postl et al. However, larger holes that gradually heal over time, and that have been imaged at various stages of the healing process, cannot be explained simply by holes diffusing away. We must have a supply of carbon atoms coming from somewhere that is driving the healing process.

As a last example, consider the work of van Dorp et al.[20] In this work, the authors demonstrate molecule-by-molecule deposition onto a graphene substrate using e-beam-induced deposition (EBID), where a precursor molecular gas is purposely supplied within the STEM. Of interest to us here is their comment, in the supplemental materials of that work, regarding their efforts to obtain clean graphene, where they say, "*While this is generally sufficient for imaging samples* *without noticable [sic] contamination, it is insufficient for large-area freestanding single-layer graphene produced with the CVD method. For single-layer graphene exhaustive additional cleaning does not reduce the contamination level;*"

Despite exhaustive precautions, they ultimately had to abandon working with chemical vapor deposition (CVD)-grown single-layer graphene and instead work with smaller multilayer graphene flakes. They conjecture that the small size and step edges on the multilayer flakes is responsible for limiting the diffusion of hydrocarbons.

For graphene, which is composted of a single layer of carbon atoms, the tolerance for deposition is much more stringent than for most other materials. Any amount of surface contamination immediately obscures the view of the sample in (S)TEM and graphene is notoriously dirty.

In spite of the challenges in obtaining and maintaining clean graphene, it has become the material of choice for investigations into atomic manipulation and fabrication using an e-beam. Such investigations began with observations of impurity atoms in graphene[21,22] (usually silicon) but advanced to directed movement,[23–25] deliberate insertion,[25] assembly,[26] and the extension to other atomic species.[27–31] A similar insertion process has been used to place a single carbon atom into a vacancy in h-BN,[32] indicating the possible extension to other 2D materials. Recent results have shown atom-by-atom writing,[33] the creation of atomic arrays,[34] and the reconceptualization of the STEM itself as a synthesis platform.[35,36]

Clearly, the requirements for cleanliness when attempting to manipulate single atoms is far more stringent than for imaging. Manipulating single atoms involves making and breaking chemical bonds. This means that one will invariably have chemically reactive structures that are prepared to bond with the next available atom. In the words of van Dorp et al.[20] "*The defects that are created by the high-energy electrons act essentially as contamination traps.*"

In this work, we would like to challenge the conceptual picture of "clean" graphene produced by *in situ* heating. Specifically, we show that there can still exist a layer of rapidly diffusing hydrocarbons loosely adhered to the sample surface. Within this picture, we also establish a few further observations: (1) these hydrocarbons can be impeded by e-beam deposits, (2) these surface hydrocarbons can be directly detected in the image intensity when their concentration is high enough, (3) the hydrocarbons desorb so that their concentration on the sample surface decreases over time, and (4) that e-beam mediated hydrocarbon deposition can occur indirectly.

This conceptual framework is critical for understanding the atomic processes observed on graphene.

**Methods**

Atmospheric pressure chemical vapor deposition (AP-CVD) grown graphene[37] was transferred to a Protochips heater chip using the graphene-water membrane transfer technique.[38] Prior to loading, the sample, cartridge, and microscope magazine were baked under vacuum at 160 °C for eight hours. A Nion UltraSTEM U200 was used for imaging, operating at an accelerating voltage of 100 kV, a nominal beam current of 66 pA, and a convergence angle of 30 mrad. The medium angle annular dark field (MAADF) detector was used for imaging, with a nominal inner angle of 60 mrad. High angle annular dark field (HAADF) detectors, with inner angles greater than about 80 mrad begin to give image contrast that approximates the square of the atomic number of the element being imaged.[39,40] Thus, heavier elements appear brighter allowing direct interpretation of image contrast. Contrast in MAADF imaging is similar to that of HAADF imaging but, due to the lower inner angle, it gives higher intensity which is important for us here. Prior to imaging the sample in the STEM, the substrate was heated *in situ* to 1200 °C to evaporate surface contaminants[16] and then reduced to 800 °C and held at this temperature for the duration of the experiments.

**Results and Discussion**

*Direct Imaging of Diffusing Hydrocarbons on Graphene*

Figure 1(a) shows an initial overview image of the graphene sample held at 800 °C. We observe the typical dendritic surface residues characteristic of graphene cleaned by *in situ* heating. Following a previously described strategy for preventing the ingress of diffusing hydrocarbons on graphene,[41] a continuous deposit of carbonaceous material was created, using manual positioning of the e-beam, to enclose a portion of the graphene. This deposited material is marked by the dashed line in the image shown in Figure 1(b). Note that material deposited in this way collects on both sides of the membrane.[42] The image intensity reveals a distinct contrast difference between the graphene inside the enclosed region and the graphene outside the enclosed region, which we call the 'open' region. The graphene outside the enclosure appears distinctly brighter. This contrast difference is not observable in the overview image and is attributed to the presence of rapidly diffusing hydrocarbons loosely adhered to the graphene surface. Such hydrocarbons are generally assumed to not be directly imageable until they become bonded to the surface via e-beam deposition, at which point they begin to obscure the images. For example, the perceptive reader may notice the small square feature on the left side of Figure 1(b) which is not present in the overview image shown in Figure 1(a). This is hydrocarbon deposition that was inadvertently deposited on the graphene surface while attempting to focus the microscope. The square shape is a result of the scan area used.

A second image of the same region was acquired 11.5 mins later, shown in Figure 1(c), where the contrast difference between the enclosed and open regions has vanished. These contrast differences are somewhat subtle to detect by eye without adjusting the brightness and contrast to such an extent as to clip other portions of the image. We have chosen to display the images without clipping to maximize their overall interpretability with regard to the various features present. To clearly illustrate the intensities in each region we provide a quantitative image intensity comparison in Figure 1(d). Here, the graphene intensity for the open and enclosed regions is compared between the image in (b), labeled 'I1', and the image in (c), labeled 'I2'. For this analysis, the upper portions of the images were used (to avoid unnecessary complications in categorizing and segmenting the bilayer regions and the support membrane) and are displayed as overlays in (d). These image clips were then segmented using the Trainable Weka Segmentation plugin[43] for Fiji[44] to discriminate between the graphene and the immobile surface contaminants (the bright regions). The pixel intensity distribution for each region is summarized in the violin plot which displays the distribution mean (white dot), one and two standard deviations (dark gray lines), and the kernel density estimate (colored) mirrored across the vertical axis. The dashed, horizontal line marks the mean of the open region of I1 for easy visual comparison to the other distributions. The open region in I1 is clearly brighter by about one standard deviation.

To determine whether this is due to the presence of diffusing surface hydrocarbons, an image sequence (50 frames) was acquired across the boundary indicated by the dashed box in Figure 1(c). Acquisition of this image sequence required approximately 37 seconds. For the rest of the time (between the frames shown in Figure 1(b) and (c)) the beam was left scanning across the larger field of view. A montage showing six representative images is shown in Figure 1(e), artificially colored using the 'Fire' look up table in Fiji. In the open region, hydrocarbon deposition occurs much more quickly indicating a higher concentration.

A previous investigation into the use of physical barriers for blocking hydrocarbon diffusion on graphene found that (under the conditions used) they were highly effective.[41] In that work, no deposition was observed within the enclosed region after encircling barriers had been deposited. Additionally, there was no detectable delay between the completion of the barrier and the suspension of deposition. Here, we observe deposition both within the enclosed region as well as outside. If we operate under the assumption that the same mechanisms underpin both results— that is, after the completion of the barrier the diffusing hydrocarbons that were originally within the enclosed region play a negligible further role in the deposition process—then we must conclude that hydrocarbons from the open region are able to diffuse across the barrier here. Under these conditions the deposited barrier has only suppressed and not completely prevented the continued ingress of diffusing hydrocarbons. It appears likely that the concentration of diffusing hydrocarbons in the open region plays a role in the effectiveness of the barrier.

Two points are worth stressing: (1) although the image contrast difference has vanished between the enclosed and open regions in image I2, this is due to detector sensitivity and noise level for these imaging conditions, and (2) the hydrocarbon concentration in the open region has dropped significantly while the sample temperature has remained constant, indicating a finite desorption time. The first point is important because we want to be careful not to suggest that the surface concentration of hydrocarbons in the enclosed and open regions equalizes after a few minutes. That is to say, just because we cannot *detect* an intensity difference between the open and enclosed regions does not necessarily imply that there is no concentration difference. In order to directly detect a difference in the image intensity, the concentrations must first have a high enough signal to detect. In Supplemental Figure S1 we show an example of deposition in an open region (and none in the adjacent enclosed region) where the graphene intensity shows no difference in hydrocarbon concentration between these regions. But the difference in deposition indicates that there must be a difference in concentration. Thus, we conclude the hydrocarbon concentration is below our detectability limit for those imaging conditions.

The second point is important because it is easy to assume, when performing STEM observations across time, that the sample state has remained constant, especially when no direct evidence of a change in the sample can be observed in the images.

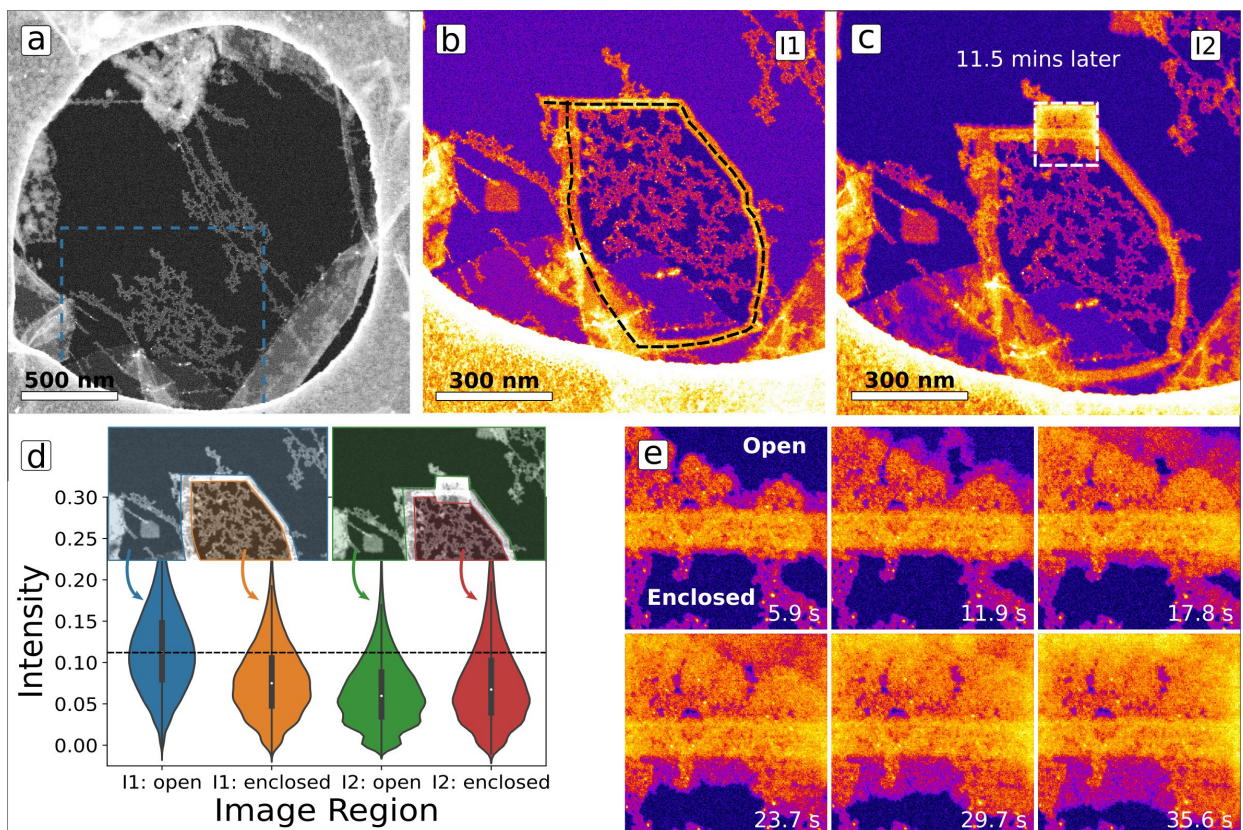

**Figure 1 Direct observation of diffusing hydrocarbon intensity.** (a) Overview image of graphene suspended across aperture. Dashed box shows approximate location of where the images in (b) and (c) were acquired. (b) Image of e-beam deposited barrier, tinted yellow. This image is labeled 'I1'. (c) Image of the same region shown in (b), acquired 11.5 minutes later. This image is labeled 'I2'. The dashed box shows the region where the image sequence summarized in (e) was acquired. (d) Violin plot of graphene intensity. The dotted line marks the mean intensity of the open region in 'I1'. (e) Montage of a 50 frame stack of images acquired across the deposited barrier indicated by the dotted box in (c).

*Further Examples of Direct Visualization of Diffusing Hydrocarbons and Their Behavior*

We have presented evidence for direct detection of rapidly diffusing hydrocarbons on graphene. In addition, we claim that an e-beam deposited barrier can substantially reduce the concentration of these hydrocarbons. However, a single example is somewhat scanty evidence. Here, we provide additional evidence of a similar nature to bolster confidence in our claims, after which we will begin to discuss a more nuanced picture of the diffusion/deposition process.

Figure 2(a) shows a second example of direct detection of diffusing hydrocarbons on a graphene surface. These hydrocarbons appear as a uniform, delocalized background intensity and are therefore difficult to detect without a reference intensity to act as the comparison. In this example, we first deposited a barrier in a square using the e-beam. This barrier is highlighted (and labeled) by the dashed square in the first panel of Figure 2(a). Subsequent imaging of the barrier produced the initial deposition along the growth front that is also labeled. The panels shown are taken from an image sequence (video) that is presented in the supplemental information. We observe hydrocarbon deposition and growth along the outside of the barrier and can see through visual inspection that the intensity outside the barrier is higher than the intensity inside the barrier. This is indicative of the presence of additional material outside the barrier.

In Figure 2(b) we show a larger field of view of the sample. The green square in the lower right corner indicates where the image sequence shown in (a) was acquired. We note that the extended region of graphene above this area is of uniform intensity. We then deposited two barriers across this region, shown in Figure 2(c). At the completion of each barrier, the intensity of the enclosed region dropped distinctly to match that of the previously enclosed region. This drop indicates a decrease in the concentration of the diffusing hydrocarbons.

In Figure 2(d) we show a magnified view of the region within the blue box in (b). Closer examination of this region reveals several sharp intensity changes. Initially, this sudden change in intensity was thought to be spurious. However, the intensity changes observed here seem to match the difference in intensity observed between the open and enclosed areas of graphene. Thus, we remain open to the possibility of an alternate explanation which we now consider.

In Figure 2(e) we again expand the field of view. The patch of single layer graphene shown in (a)-(d) can be seen in the lower right corner. The rest of the field of view consists of mostly bilayer and some patches of three or more layers of graphene. In this region, we again performed a similar experiment, enclosing a square region using e-beam deposition. Again, on further imaging hydrocarbon deposition and growth was observed only outside the enclosed region. We can see in the image a clear difference in intensity between the enclosed bilayer and the open bilayer regions. Before moving to the next panel, we note that the intensity in the open region is uniform.

In Figure 2(f) we show an image where we increased the gain on the detector to the point that many of the brighter regions are saturated. However, this gives us a greater sensitivity to the fluctuations in the intensity of the faint hydrocarbons that are of interest here. During the image acquisition we observed a sudden decrease in intensity in the open bilayer regions of the sample. This sort of intensity change can be due to changes in the beam current or detector gain, so is normally ignored. However, if this feature were an image artifact it would be found across the entire width of the image. To rule out the possibility that this is an image artifact we can check whether it is found consistently across the image horizontally. We see that in the single layer area the sharp decrease in intensity is not present. There does appear to be a small intensity gradient on the single layer graphene in this area and at the magnification shown here it is difficult to see this region clearly enough to make a conclusive decision. In the supplemental materials we show magnified views of each of these regions where there is a sharp intensity change and compare these with the single layer area. We use horizontal lines as reference markers and conclude that while there is a very small intensity gradient in the single layer area, it is not sharp like the other regions and does not appear in the correct location to be consistent with the interpretation that it is an image artifact. We conclude that the intensity change is local and thus is a real change in the sample. We use the short-hand 'no intensity change' in the figure to mean that this sharp feature is not present and in the correct location for it to be interpreted as an image artifact.

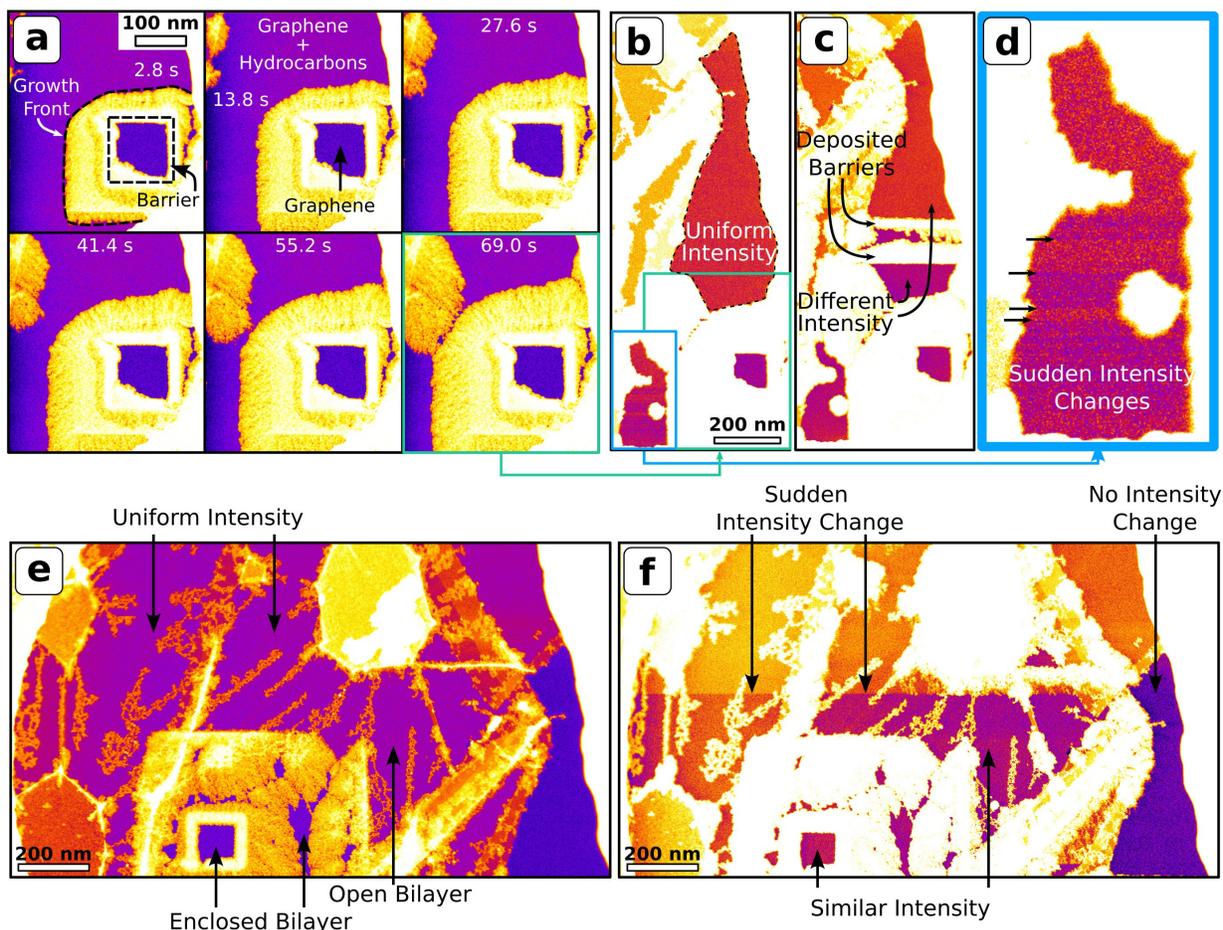

**Figure 2 Further examples of the detection of diffusing hydrocarbons.** (a) Montage showing six frames from a video of hydrocarbon deposition and growth around an e-beam deposited barrier on graphene. (b) Larger field of view. The green box represents the region examined in (a). We observe a uniform intensity within the dotted region. (c) Two barriers were deposited across the dotted region resulting in a decreased intensity in the enclosed region that is directly visible in the image. (d) Magnified view of the blue boxed region in (b). We observe conspicuous and sudden intensity changes during the image acquisition. (e) In a bilayer region of the sample a similar effect was observed, a square barrier was deposited, after which deposition only occurred in the open region. We note the contrast difference between the open bilayer and enclosed bilayer regions. We also note that the open region has a uniform intensity. (f) Example of a sudden and pronounced intensity change. This change does not occur across the entire image.

A quantitative comparison of the image intensity values observed in the images shown in Figure 2(b) and (f) are shown in Figure 3(a) and (b), respectively. Figure 3(a) shows a density plot comparison between the intensity of the enclosed graphene, labeled 'Grap.', the intensity of the graphene and hydrocarbons, labeled 'Grap. + Hydro.', and the upper region of the area shown in Figure 2(d), labeled 'Top Region'. (The Top Region is plotted as a line instead of a histogram to reduce visual clutter.) The mean of each of these distributions is marked by the vertical dashed lines. An intensity profile across the region with sharp intensity variations was extracted from the image in the location marked by the purple rectangle and the line profile is overlaid vertically on the density plot so that the profile intensities correspond with the x axis intensities, labeled 'Intensity Profile'. In this way we can directly compare the line profile intensities to the intensities of the reference areas. The sharp changes in intensity are marked by arrows. The quantitative change in intensity approximately corresponds to the difference between clean graphene and graphene with the diffusing hydrocarbons.

A similar analysis is shown in Figure 3(b). The intensity of a region of enclosed graphene is labeled 'Encl. Grap.'; the intensity of a region of open graphene that was formerly brighter is labeled 'Open Grap. 1'; the intensity of a region of open graphene that is bright, indicating graphene and diffusing hydrocarbons, is labeled 'Open Grap. 2'; a line profile was taken across the boxed region, where we observe a sharp intensity change, and is overlaid on the density plot. Here, again, we see that the change in intensity is approximately the same as that between clean (bilayer) graphene and (bilayer) graphene plus diffusing hydrocarbons. The sharp intensity change is marked by the straight arrow. The large spike in intensity labeled 'Contamination' is due to a bit of surface contamination in the region where the profile was taken. The Python code that reproduces this analysis is available at https://github.com/ondrejdyck/Your-Clean-Graphene-is-Still-Not-Clean.

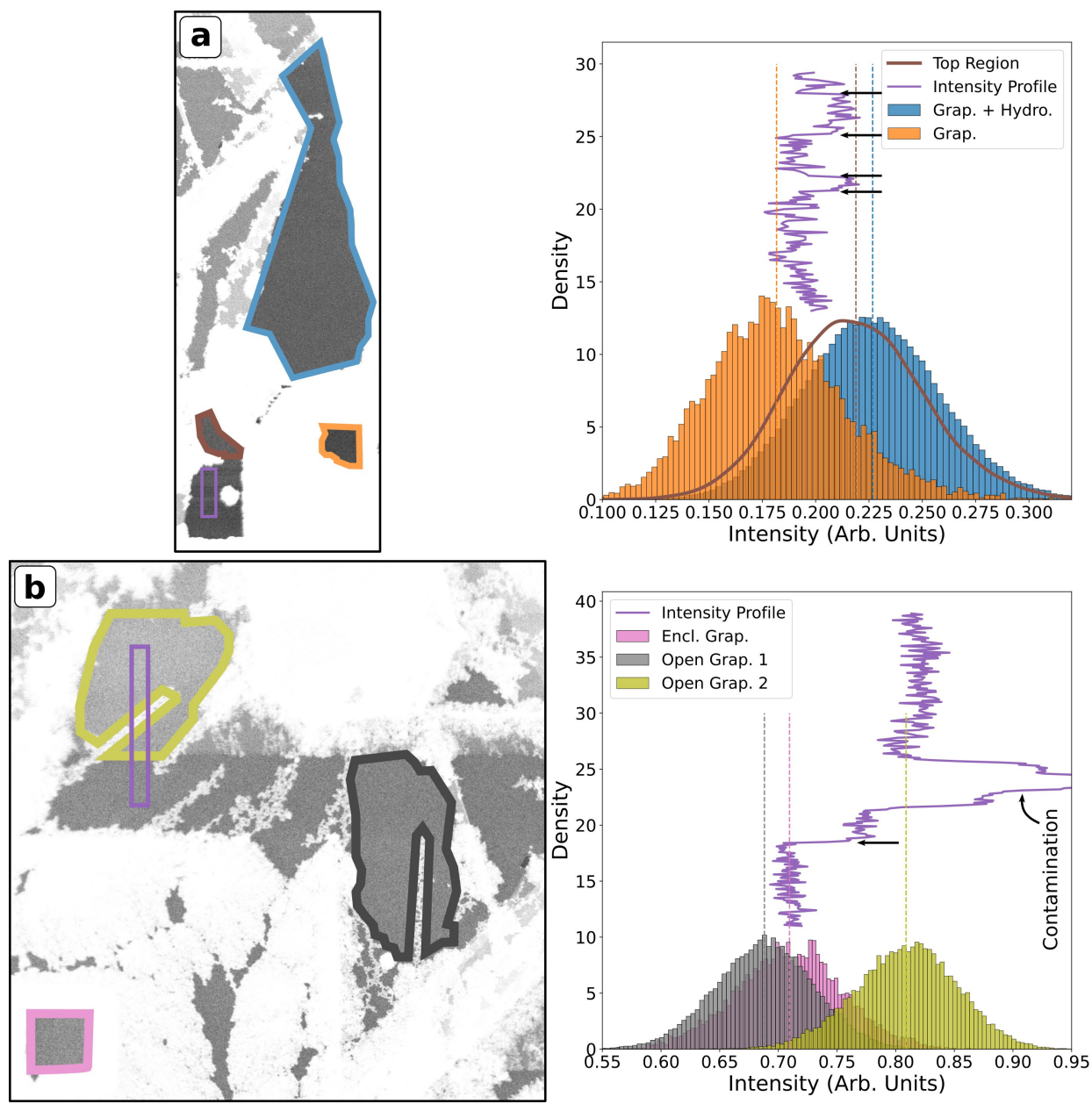

**Figure 3 Quantitative comparison of intensities.** (a) Quantitative comparison of intensities observed in Figure 2 (b). The boxed patches in the image are color coordinated with the histogram. (b) Quantitative comparison of intensities observed in Figure 2 (f).

The only hypothesis we can think of, that may explain the sharp changes in observed intensity, is a rapid change in the concentration of surface hydrocarbons. Previous observations have indicated that these hydrocarbons desorb over time,[45] which is consistent with the observations shown in Figure 1(b) and (c), where the intensity of the unenclosed graphene has measurably decreased over several minutes. The abrupt change is more surprising. This observation is likely related to observations that as soon as an area becomes enclosed, the hydrocarbon availability decreases faster than we have been able to detect thus far. We have not observed an enclosed area that begins with a high intensity (high concentration of diffusing hydrocarbons) and gradually decreases in intensity.

When considering potential mechanisms, the time scale over which this decrease should be expected will depend on both the diffusion rate of the hydrocarbons and their concentration. A high diffusion rate can compensate for a high concentration and vice versa. Once an area becomes enclosed, a low concentration of hydrocarbons could be expended very quickly through e-beam deposition, resulting in a sharp decrease in observed intensity. If this is the correct interpretation then there must exist some channel through which hydrocarbons are being supplied to the imaged regions and this channel must become blocked (and unblocked in the case shown in Figure 2(d) where the intensity both decreases and increases) leading to a rapid change in intensity within the now enclosed region.

Alternatively, it is possible that spontaneous desorption occurs at a more rapid rate than what we have assumed, or what is apparent. In Figure 1(b) and (c) we observed a marked decrease in hydrocarbon concentration, manifesting in a decreased image intensity, over the course of ~10 min. Previous investigations[45] (see supplemental Figure S3 in that work) documented a decreased e-beam induced hydrocarbon deposition rate through time at 900 °C over the course of 80 min. These observations suggest a gradual desorption over the course of minutes to hours. However, it is possible that rapid desorption is occurring and at the same time being compensated by a continual resupply of hydrocarbons evaporating from an unidentified source(s). As the source evaporates it frees hydrocarbons to migrate along the sample surface for a time before they are desorbed. These two competing processes, desorption and resupply, would create a situation where the apparent desorption time appears much longer than the actual desorption time. Once the source is completely expended or the channel that allows mass transport to the observed region is blocked, we would expect to observe a sudden decrease in hydrocarbon concentration due to desorption because the mechanism for resupply is stopped.

These hypotheses are, at this point, conjectural and we do not claim that we have established their validity beyond doubt. However, we believe that the inclusion of these observations here may help to stimulate curiosity and further investigations.

Let us return to Figure 2(a) and consider the dynamic evolution that is observed. Most electron microscopists will recognize this kind of behavior as a heavily contaminating specimen. Aside from expressing annoyance at its obtrusive character in covering the intended sample, little thought is put toward an elaboration of what is being observed and how this process proceeds (at least, this has been our own attitude toward such observations in the past). We clearly observe a growth front where hydrocarbons are collecting and becoming fixed in place. Borrowing from the standard interpretation of e-beam-induced deposition (EBID) the precursor material, here hydrocarbons, are being dissociated or broken apart through an interaction with the e-beam. Once they have been dissociated they become chemically reactive and bond with the substrate. However, this interpretation does not fit the observations shown in Figure 2(a). Here, we are scanning the e-beam across the whole field of view repeatedly to acquire the image sequence. But we do not observe deposition across the whole field of view, we observe deposition only along the growth front. It is, of course, possible to obtain deposition directly under the e-beam under certain conditions but to interpret the images shown in Figure 2(a) we must add another element to the EBID interpretation, namely, a diffusion of the reactive dissociated hydrocarbon fragments to a bonding location. These hydrocarbon fragments must be generated across the whole field of view where we can, in the image contrast, directly see the presence of the hydrocarbon background intensity. There is no reason to suppose that the dissociation process only occurs along the growth front. We could suppose that an increase in secondary electrons, produced at the growth front, more strongly interact with the hydrocarbons thus producing a preponderance of reactive fragments primarily at the growth front. Nevertheless, this interpretation does not forbid or explain the often-observed nucleation of deposit sites in a pristine area. Thus, we must conclude that the diffusing hydrocarbons are being dissociated across the totality of the irradiated region (where the hydrocarbons are present). But we do not observe deposition across all of this region, we only observe deposition at the growth front. To explain this observation we infer that dissociated and chemically reactive hydrocarbons do not immediately bond at the location where they are dissociated, but also diffuse away from their point of generation and attach to the nearest location where the hydrocarbon fragment finds a strong chemical bond that is able to hold it in place against diffusion.

If this description is accurate, we should be able to arrange a situation where we position the beam at position, A, but observe deposition at some other location, B. This would conclusively demonstrate that the dissociated hydrocarbons do not necessarily bond directly to the substrate at the point of generation, but first diffuse along the sample surface until they find a suitable bonding location. The following experiment illustrates this possibility.

*Indirect E-beam Deposition*

The typical conceptual description of EBID, as mentioned above, follows along these lines: (1) a gas phase source material—here, hydrocarbons—are delivered and flow/diffuse randomly across the sample surface, (2) the energetic primary e-beam imparts enough energy to these diffusing molecules to dissociate them, i.e. to break them apart, leaving chemically reactive molecular fragments, (3) these fragments quickly bond to the sample surface building up a material deposit at the e-beam location. This simplified description neglects diffusion of the reactive chemical fragments, generated in step two, from their point of origin to the final bonding location which forms the deposition site conceptualized in step three.

This model works acceptably well in most cases and adheres nicely to the principle that everything should be made as simple as possible and not simpler. However, as we have already highlighted in the introduction, atomic-scale fabrication requires a detailed understanding of the chemical processes involved at the level of single atoms and molecules. As we venture into this field it is important that we attend to the finer details of our mental models which we use to furnish our thoughts with simplified abstractions of the real world. We implicitly assume that such abstractions, however simplified, capture the *relevant* behavior of the system. Therefore, as one ventures further into unexplored territory, it is prudent to occasionally check that our simplified descriptions still accord with the relevant behavior. As we have seen in the previous discussion, we run into difficulty applying this simplified description to the observations shown in Figure 2(a) and have added another diffusion process, although it is not directly observed, to explain why hydrocarbon growth does not occur across the whole field of view.

As a way of conclusively demonstrating this phenomenon we posit that we should be able to indirectly deposit hydrocarbons at a region adjacent to the irradiated region without observing any deposition under the e-beam. This will rule out the possibility, mentioned above, that secondary electrons generated at the growth front are responsible for all of the dissociation. In fact, we have no grounds here to suggest that *any* dissociation stems from these secondary electrons but the mere fact that it is possible dictates that we must allow for it in our attempts to assert that we expect dissociated fragments to exist across the entire irradiated area. It is this postulate, which is not being directly observed, that leads us to infer diffusion of the dissociated hydrocarbons, which is also not directly observed.

Thus, to solidify this argument, we show a case of purely indirect deposition, shown in Figure 4. Figure 4(a) shows the initial sample state. A small subscan box was positioned at the location of the red square and a series of images acquired while scanning this region of the sample. The inset image shows an example subscan frame. With these images we were able to verify in real time that no deposition occurred during the exposure. After delivering a dose of $4\times10^{10}$ electrons at the location of the red square, the base image shown in Figure 4(b) was acquired. To highlight the differences, the image shown in (a) was tinted blue and overlaid on the base image in (b). Regions that exhibited hydrocarbon deposition protrude beyond the blue tinted overlay and appear white.

We draw two main insights from this experiment. The first is that, although we can acquire atomically resolved images of the lattice atoms (e.g. see the inset in Figure 4(a)), these image features do not deliver all of the information about the sample state. This is visually represented in Figure 4(c) where we provide a conceptual rendering of atomically clean graphene vs. graphene encompassed by a cloud of rapidly diffusing hydrocarbons. The directly interpretable image features communicate only the concept of atomically clean graphene, while this experiment demonstrates that there also exists a significant amount of 'invisible' hydrocarbons. Given sufficiently high concentrations, these hydrocarbons are directly detectable in the image intensity as shown in Figure 1 and 2, but even in cases where the image intensity does not appear to show a higher hydrocarbon concentration, they may still be present in non-negligible amounts (e.g. consider the number of atoms required to build up the deposits shown in Figure 4(b)).

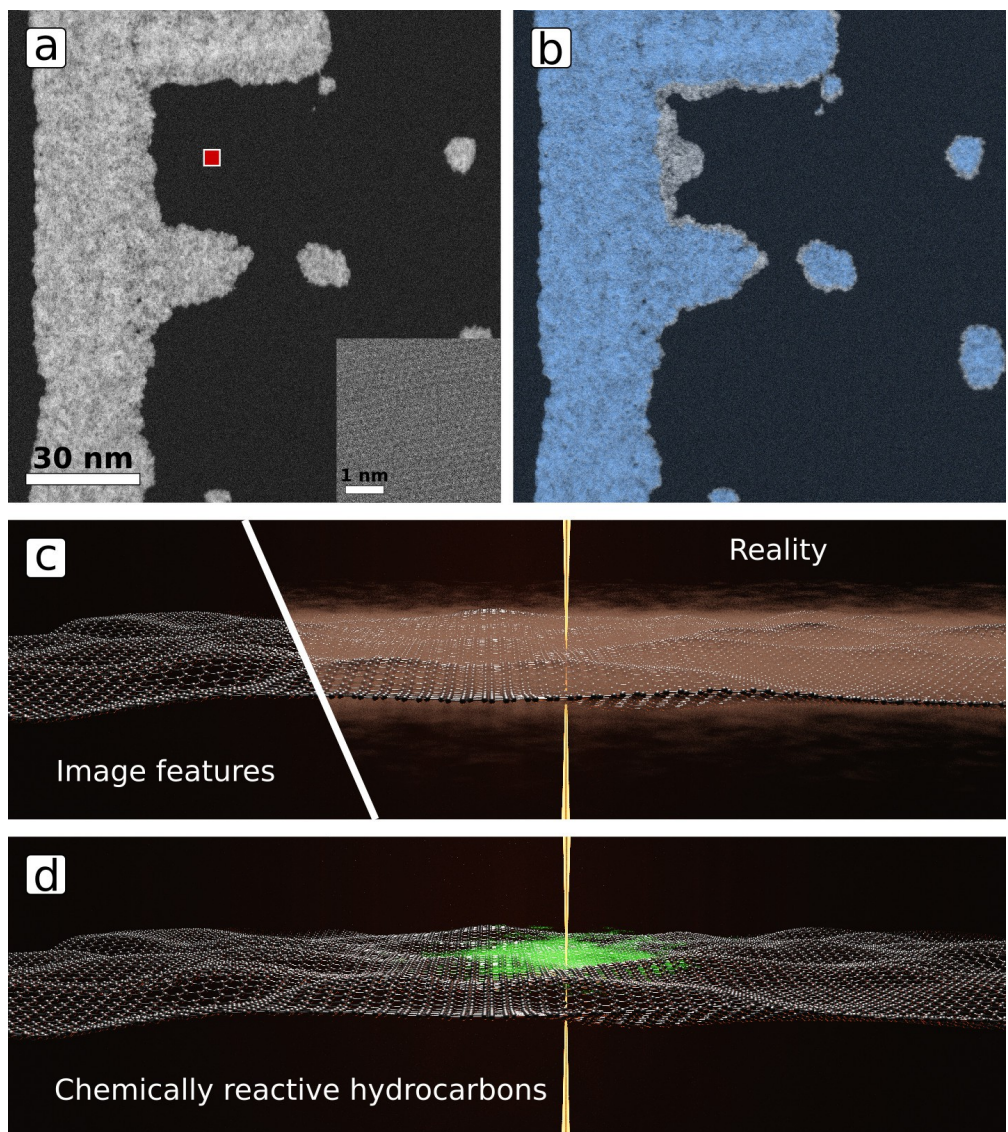

**Figure 4 Indirect e-beam deposition.** (a) Initial sample configuration. E-beam was scanned in the area indicated by the small red box. An example image of the subscanned area is shown inset. After approximately 95 seconds of exposure to the subscan region, the base image show in (b) was acquired. The blue overlay is a semi-transparent version of (a). The white regions protruding beyond the blue areas are indirectly deposited hydrocarbons. (c) Concept rendering of graphene surrounded by a cloud of rapidly diffusing hydrocarbons, labeled 'Reality', compared with the primary image information source, labeled 'Image features'. (d) Concept rendering of reactive hydrocarbons diffusing away from the e-beam. The rest of the unreactive hydrocarbons are not shown to reduce visual clutter.

The second insight is that chemically reactive hydrocarbons that have been dissociated by the e-beam do not necessarily chemically bond to the surface of graphene at the e-beam location. Instead, these reactive hydrocarbons continue to diffuse away from the e-beam location as depicted in Figure 4(d). In this depiction, the unreactive hydrocarbons have been omitted from the rendering to reduce visual confusion. The reactive hydrocarbon fragments are depicted as a green gas diffusing along the graphene surface away from their point of origin, the e-beam. This experiment was attempted several times before this result was obtained, each time eventually resulting in deposition at the e-beam location. We stress that direct e-beam deposition at the beam position is obviously possible and likely, but diffusion of the dissociated hydrocarbon fragments must also be taken into account to obtain a full picture of the process.

*Discussion*

We have shown a number of effects of diffusing hydrocarbons and their interaction with the electron beam. (1) We showed that hydrocarbon deposition can be used to create physical diffusion barriers that limits the ingress of additional hydrocarbons.[45] (2) We showed that sufficiently high concentrations of diffusing hydrocarbons can be directly detected in the image intensity and that the difference in hydrocarbon concentration between enclosed and open regions of graphene can be seen in the images. (3) We showed that sudden changes in image intensity suggest that hydrocarbon concentration can dynamically vary. A full understanding of the origins of this variation would be a significant step toward the ability to control surface contaminants. (4) We showed that e-beam-induced hydrocarbon deposition involves a diffusion step: After the source hydrocarbons are dissociated, they must also diffuse to the location where they will bond with the substrate. The final bonding location is not necessarily the location where they were dissociated. While this effect seems to be logically intuitive, by showing purely indirect e-beam deposition, we illustrate that it is also significant—at least for graphene.

In light of these results, we turn to discuss the often-used strategy of a 'beam shower'.[46] This procedure involves illuminating a large area with the electron beam for a few minutes (~10-30 mins) for the purpose of preventing hydrocarbon deposition during subsequent imaging. The physical explanation that is usually advanced is that the beam shower dissociates the majority of the diffusing hydrocarbons within the irradiated region, pinning them to the sample surface, and leading to a depletion of free hydrocarbons. After they have been depleted, high resolution imaging can be performed without further deposition within the previously irradiated region. The effectiveness of this strategy is temporary. With seeming inevitability, the deposition eventually returns and another beam shower will be required.

In contrast to this picture of inevitably returning hydrocarbons, our previous work showed that e-beam deposited corrals, like that shown in Figure 1, could permanently (days) suppress the ingress of hydrocarbons.[45] The findings of this work temper our previous optimism by showing that with a high enough concentration of hydrocarbons the barrier can be surmounted, leading to deposition within the corraled/enclosed region. However, these findings suggest another mechanism that may be operating during the beam shower process.

Frequently, a beam shower will produce much thicker deposition at the edges of the irradiated region.[46] This deposition may be inadvertently performing the same function as the intentionally deposited barriers shown in this work. If this is the case, one would expect to observe not simply a temporary alleviation of deposition but, even after the deposition returns, a permanent reduction in the deposition rate. To our knowledge this effect has not been systematically analyzed. It is also possible that various competing phenomena might complicate the quantitative evaluation at high contamination levels.

While the description advanced in this work more fully accounts for the observations, it should be mentioned that there are still many unknowns. Many species of hydrocarbon exist on graphene surfaces.[47] The diffusion constant of these hydrocarbons is unknown. The concentration of these hydrocarbons is unknown. Very likely the different species will affect the concentrations and diffusion constants so that there is not just one diffusion constant and concentration but as many diffusion constants and concentrations as there are species of hydrocarbons. Likewise, the same considerations can be applied to the dissociated fragments. Each fragment may have a different diffusion constant and the different modes of dissociation may produce many chemically distinct fragments, each with its own properties. It is also possible, and likely, that all these parameters vary from sample to sample. The degree to which an average hydrocarbon response can be established that can be relied upon to describe and predict the behavior of hydrocarbons on samples in general, or hydrocarbons on graphene samples in particular is also unknown.

There is ample room for an enhancement of our understanding along these lines and the work presented here represents a determined step in the direction of a more detailed description of these interactions. We hope that this work will not only establish a more nuanced view of hydrocarbons on graphene but also help draw interest toward a deeper scientific understanding of these processes. Future efforts toward atomic scale fabrication and the technologies that will be built upon such capabilities will invariably depend on our ability to exercise an exquisite degree of control over the sample environment.

**Conclusion**

In this work, we have examined the subtle and somewhat unexpected behavior of hydrocarbons on the surface of graphene. We have shown that it is possible to directly detect rapidly diffusing hydrocarbons on graphene. These hydrocarbons are not directly imaged due to their rapid motion. Nevertheless, their presence can be detected as a uniform background intensity. These hydrocarbons were observed to decrease in concentration over time, which is attributed to a previously reported desorption process, and their detectability, thus, also decreases with time. The creation of physical barriers can reduce or completely suppress the availability of hydrocarbons and, therefore, deposition within enclosed regions. This indicates that the layer of diffusing hydrocarbons must be (loosely) bound to the graphene surface and does not extend arbitrarily into the vacuum. If this were not the case, we would see deposition everywhere including within the enclosed regions. Finally, we give conclusive evidence for indirect deposition which indicates that chemically reactive hydrocarbon fragments that are generated at the e-beam position diffuse across the sample and bond to locations elsewhere. Taken together, these results present an updated picture of the dynamics of hydrocarbons on what otherwise appears to be clean graphene. These results represent another step in understanding and control over atomic scale cleanliness and beam-sample interactions that form the fundamentals of atomic manipulation and fabrication.

**Acknowledgements**

This work was supported by the U.S. Department of Energy, Office of Science, Basic Energy Sciences, Materials Sciences and Engineering Division (O.D., A.R.L, S.J., K.X.) and was performed at the Oak Ridge National Laboratory's Center for Nanophase Materials Sciences (CNMS), a U.S. Department of Energy, Office of Science User Facility. A.O. and S.L. were supported by NSF DMR-2105126.

**Supplemental Information**

In the main text, we showed examples of e-beam-induced hydrocarbon deposition in regions of the sample where the intensity of the graphene prior to deposition was higher than that of enclosed areas where no deposition was observed. In Figure S1 we show a comparison of two regions of graphene that have the same observed intensity but deposition only occurs in one region. The purpose of this example is to show that the background intensity of the diffusing hydrocarbons is not necessarily detectable even though it is present.

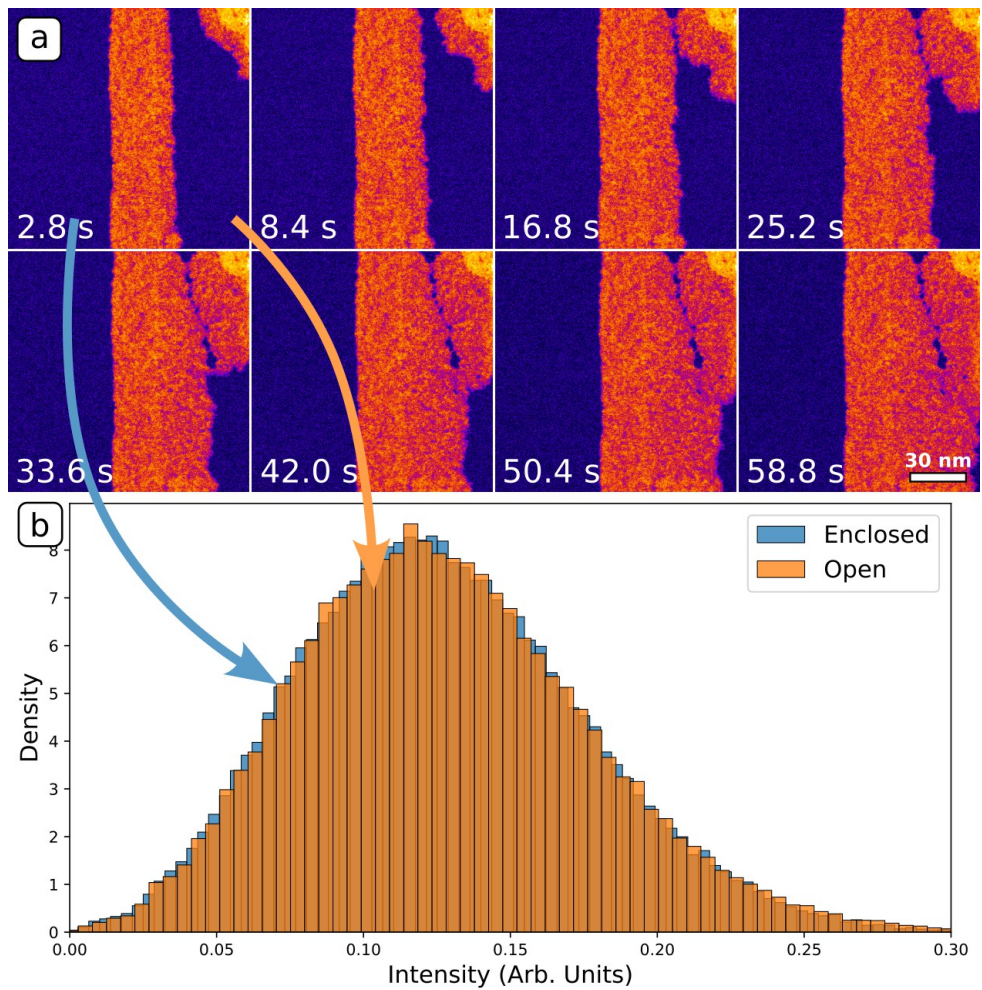

**Figure S1 Intensity comparison between graphene and graphene with hydrocarbons.** (a) Image montage of deposition occurring during e-beam irradiation. (b) Intensity comparison of the graphene on either side of the barrier in the middle. The left side is labeled 'Enclosed' and the right side is labeled 'open'.

In the main text, Figure 2(f) showed an image of a sharp change in intensity which was tentatively attributed to a change in surface hydrocarbon concentration. A significant motivation for this interpretation was the observation that the intensity change was not global. Here, we provide a closer examination of this image and the observed intensity change. Horizontal dotted lines were overlaid above and below the intensity change to provide a consistent reference mark across the image. Magnified views of the regions within the dotted boxes are provided as overlays. Significantly, the right-most box, taken across the single layer graphene area, does not show a similar sharp intensity transition as observed in the other areas.

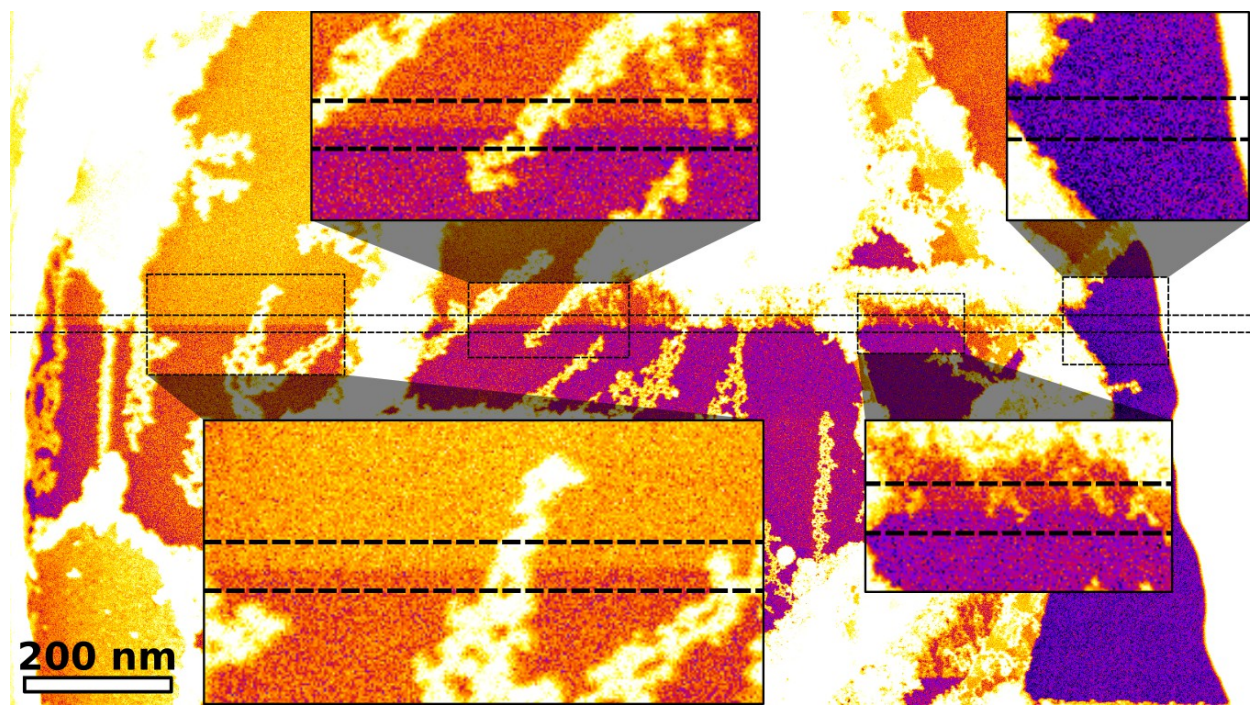

**Figure S2 Magnified view of various regions of the sharp intensity transition.**